\pdfoutput=1
\documentclass[sigconf, nonacm]{acmart}

\usepackage{multirow}

\renewcommand\footnotetextcopyrightpermission[1]{}

\AtBeginDocument{%
  }

\begin{document}

\title{BLINK: \textbf{B}atch Norma\textbf{l}ization-based \textbf{In}tegrity Chec\textbf{k}points for In-Situ Detection and Mitigation of Diverse Weight Corruptions in DNN Accelerators}

\author{Marzia Khan}
\affiliation{%
  \institution{Purdue University}
  \city{West Lafayette}
  \state{Indiana}
  \country{USA}
}
\email{khan637@purdue.edu}

\author{Akul Malhotra}
\affiliation{%
  \institution{Purdue University}
  \city{West Lafayette}
  \state{Indiana}
  \country{USA}
}
\email{malhot23@purdue.edu}

\author{Sumeet Kumar Gupta}
\affiliation{%
  \institution{Purdue University}
  \city{West Lafayette}
  \state{Indiana}
  \country{USA}
}
\email{guptask@purdue.edu}

\renewcommand{\shortauthors}{Khan et al.}

\begin{abstract}
In safety-critical deployments, AI hardware must remain reliable against a broad spectrum of threats such as radiation-induced upsets, aging,  hard faults, and adversarial attacks (e.g. progressive bit flip attack (PBFA)). All of these corrupt stored weights while the chip keeps producing confident (but inaccurate) predictions. Detecting and mitigating such weight perturbations is crucial for safety-critical platforms. To that end, we propose BLINK, an on-chip batch normalization (BN)-based on-the-fly detection and mitigation approach, which is based on continual sensing of the shift in the activation statistics, targeting a wide variety of weight corruptions (random and localized faults as well as adversarial bit flips). BLINK operates in two phases: (1) off-line pre-characterization of the relationship of the activation shifts with inference accuracy drop  (2) on-chip runtime detection and mitigation of weight corruptions. Our technique is designed so that benign perturbations stay silent while harmful faults are flagged. Upon detection, the flagged layer is re-centered to bring it closer to its stored clean reference within the same forward pass. BLINK is fully autonomous, eliminating the need for host communication, operation halts, or access to fine-tuning data. If the residual shift after mitigation indicates that accuracy has fallen below a user-set floor, a held-out watcher aborts the inference. Evaluated on ResNet-20/50 and MobileNetV2 for CIFAR-10/100, BLINK detects harmful corruptions with $>99\%$ precision across all fault types. Furthermore, it recovers accuracy from 10\% to 85.88\% under 0.5\% random bit flips (Resnet-50/CIFAR-10), up to 84\% for localized faults (MobileNetV2/CIFAR-10), and from random-guess accuracy to 80\%-83\% under PBFA (ResNet-20/CIFAR-10), entirely on chip, within the same inference pass. This comes with $<2\%$ latency cost and $0.53\%$ computation overhead.
\end{abstract}

\maketitle
\pagestyle{plain}

\section{Introduction}
Deep neural networks (DNNs) are increasingly being deployed in safety-critical applications such as autonomous driving, industrial automation and medical imaging, where incorrect predictions can lead to catastrophic consequences. Such considerations are becoming even more significant as DNN inference is steadily moving from cloud servers to edge accelerators operating under strict resource constraints. To meet the performance targets, edge accelerators often employ techniques such as aggressive voltage scaling. However, such approaches adversely impact the memory operation,  exposing weight memories to a broad range of failures.

These failures arise from several distinct mechanisms. Transient faults or gradual wear out manifest as bit flips in stored weights which produce silent data corruption. Hard faults such as stuck-at faults lead to persistent weight corruption that degrades model accuracy from the outset.  Failures may be randomly distributed or spatially localized producing structured corruption patterns (e.g., defective SRAM banks, thermal hotspots, or regional voltage droops). In addition to such environment-induced faults, weight memories are susceptible to adversarial corruption such as progressive bit-flip attacks (PBFA), in which an adversary flips a small number of high-impact bits to maximize degradation~\cite{rakin2019bit, rakin2022tbfa}, a threat made practical by injection primitives such as Rowhammer.

To mitigate the impact of weight corruptions, several approaches have been explored but with several limitations. First, a majority of them \cite{quan2022training,9643556} are designed to address a single fault class or a limited subset of failure mechanisms. Second, with a couple of exceptions \cite{ma2024dr,kundu2024mendnet}, no technique is fully autonomous and requires halting the inference upon detection of faults, require communication with a host, need access to labeled dataset, and/or rely on knowledge of per-chip fault maps. The ones that are indeed autonomous either monitor only a portion of the network for corruptions (limiting their efficacy) \cite{ma2024dr} or are designed for specific AI architectures \cite{kundu2024mendnet}.

Autonomy in AI accelerators is crucial for mission critical applications and a truly autonomous system must continuously monitor the integrity of its weights during normal operation and detect such corruptions as they occur (e.g. due to weight drifts over time, or due to a sudden adversarial attack). Once a corruption is identified, it must mitigate the error in situ. Moreover, detection/mitigation must occur without pausing inference, ensuring that no mission-critical inputs are missed, and without relying on an external host, retraining or fine-tuning data, or any other external intervention.

With this objective in mind, we design BLINK to provide full autonomy (as defined above) while remaining AI architecture-agnostic and broadly applicable to diverse weight corruption scenarios, including random and localized perturbations, PBFA, and other reliability-induced weight errors. In addition, BLINK flags harmful perturbations while remaining silent on the benign ones. For this, we use batch normalization (BN) layers as in-situ monitors of the drifts in activation distributions caused by the weight corruptions to detect harmful perturbations. If flagged, BLINK adjusts the normalization parameters to bring the activation distributions closer to the clean reference. In addition, if it detects that the faulty chip is unsalvageable, it issues an abort signal.

It is important to note that BN has been used in previous works for repair or fault tolerance, such as in Forward Parameter Tuning (FPT)~\cite{quan2022training} for stuck-at faults, BN-Fine Tuning (BN-FT)~\cite{bhattacharjee2023examining} for ReRAM-based computing-in-memory designs targeting IR-drop mitigation, and calibration of analog in-memory inference~\cite{joshi2020accurate}. However, all these techniques rely on an offline step invoked after damage is flagged and require a host and/or labeled data for mitigation.

\begin{figure*}[t]
    \centering
    \makebox[\textwidth][c]{%
        \includegraphics[
            width=1\textwidth,
        ]{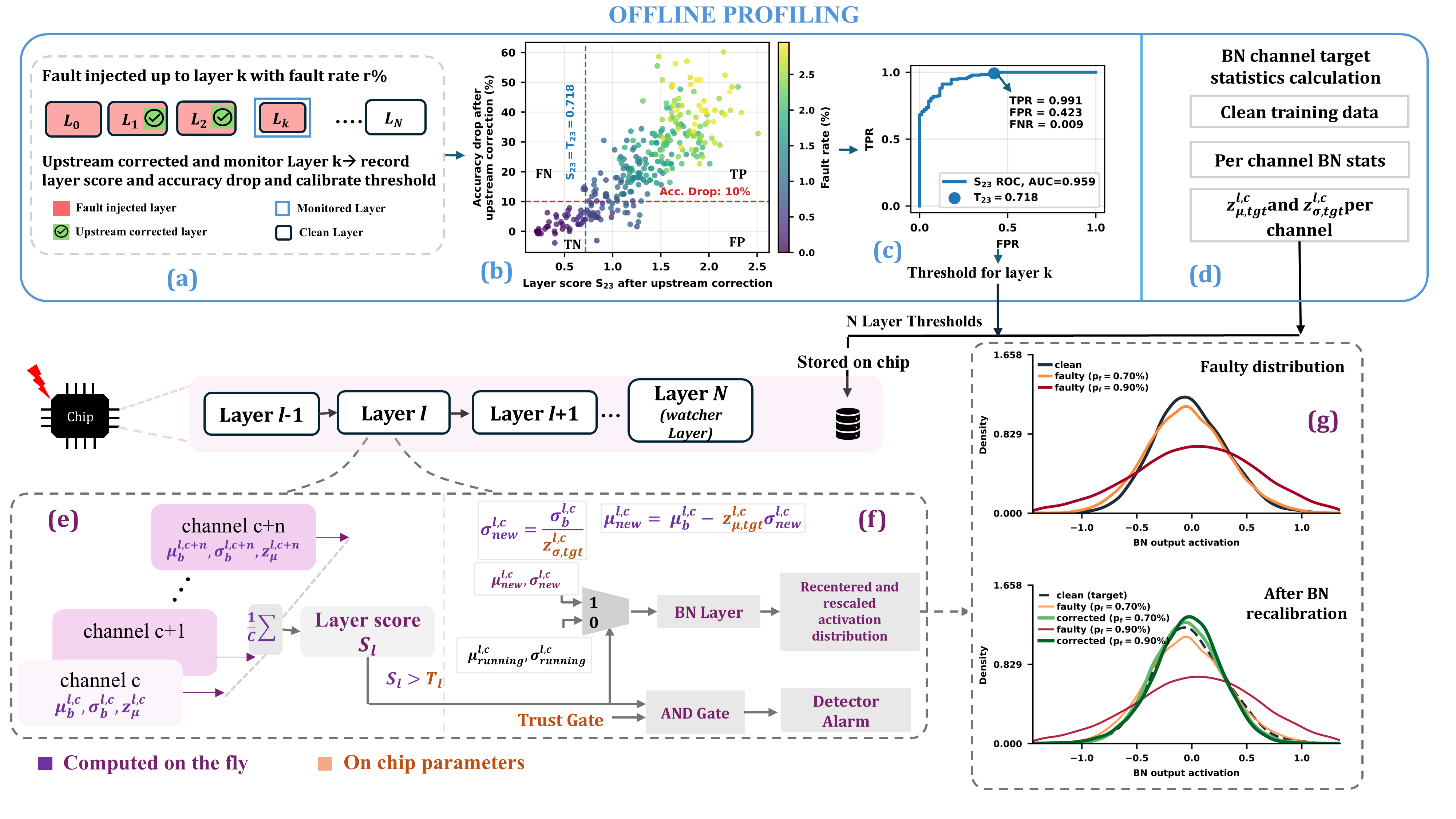}
    }
    \caption{BLINK overview. Offline: fault-injection profiling calibrates per-layer thresholds and stores clean BN targets $z_{\mu,tgt}^{l,c},z_{\sigma,tgt}^{l,c}$. Online: each BN layer scores its shift $S_\ell$; a flagged layer ($S_\ell>T_\ell$) is recalibrated in place before activations propagate.}
    \label{fig:overview}
\end{figure*}

In summary, this paper makes the following contributions:
\begin{itemize}
\item We propose BLINK, an on-the-fly and autonomous detection and mitigation technique for diverse weight corruptions.
\item BLINK enables continual in-situ monitoring of activation statistics and upon detection mitigation involves re-centering and re-scaling the activations partially nullifying the weight perturbation-induced shifts.
\item We incorporate a held-out watcher in BLINK that monitors residual shift after mitigation and triggers an abort alarm.
\item We evaluate BLINK across various weight corruption types and show excellent detection and comparable mitigation performance (with respect to previous techniques), all with full autonomy.
\item We discuss how BLINK can be combined with other techniques for further improvement in inference accuracy.
\end{itemize}

\section{Related Works}\label{related_works}
The classical approaches for fault tolerance include redundancy ~\cite{lyons1962use}, algorithm based fault tolerance (ABFT)~\cite{huang1984algorithm,zhao2021ftcnn}, and error correction codes (ECC) \cite{cojocar2019exploiting}. These methods are generally effective, but incur significant overheads, partly due to their inability to distinguish benign and harmful perturbations in the context of DNN workloads. Clipping out of range activations ~\cite{chen2021low,hoang2020ftclipact} is another technique, which suppresses the damage but is less effective for in-range perturbations. Several works have explored fault aware training~\cite{stutz2021bit,kim2018matic,he2020defending} but these methods are expensive and the hardening is only as good as the fault assumed during training.

More recently, on-the-fly and autonomous detection and mitigation methods have been explored \cite{ma2024dr}~\cite{kundu2024mendnet}. Dr.DNA\cite{ma2024dr} calculates the activation distribution signature and matches it with a profiled value for some selected portion of the network. It may not be as effective for localized faults, especially those which fall outside of these selected regions. MENDNet~\cite{kundu2024mendnet} measures energy-based uncertainty at the exits of a multi-exit network and infers from an exit that remains confident. It routes around the damage instead of repairing it and requires trained early-exit heads.

FPT~\cite{quan2022training}, BN-FT~\cite{bhattacharjee2023examining}, and analog-CIM calibration~\cite{joshi2020accurate} all use BN statistics as an offline repair mechanism. \cite{quan2022training}
and \cite{joshi2020accurate} both updates the BN running statistics with unlabeled data either \cite{quan2022training} once before deployment and or \cite{joshi2020accurate} periodically. BN-FT~\cite{bhattacharjee2023examining} instead backpropagates through the BN affine parameters, recovering more accuracy but requiring labeled data and a host.

For PBFA, defenses such as hash signatures~\cite{9643556}, encoding-based checks~\cite{liu2020concurrent}, honeypots~\cite{liu2023neuropots}, group-zeroing recovery~\cite{9474113}, host-side reconstruction~\cite{li2020defending}, and T-BFA-specific hardening~\cite{wei2024alert} have shown promise but share three limitations: (1) The protected region is fixed offline so flips outside it pass unchecked. (2) Detection methods do not distinguish between benign and harmful perturbations. (3) Recovery is either lossy~\cite{9474113} or routed through the host~\cite{li2020defending}, which under Rowhammer is the very machine running the attack.

Each of these methods commits to a fault model. \cite{lyons1962use,huang1984algorithm,zhao2021ftcnn,chen2021low,hoang2020ftclipact,ma2024dr,kundu2024mendnet} are targeted for random faults, \cite{9643556,liu2020concurrent,liu2023neuropots,9474113,li2020defending,wei2024alert,he2020defending} for bit-flip attacks, \cite{liu2017rescuing,meng2021self,quan2022training,malhotra2025weight} for stuck-at faults, and \cite{joshi2020accurate,bhattacharjee2023examining} for analog drift and device variability.

BLINK is a generalized approach covering a broad spectrum of faults with BN based detection and mitigation completely autonomously.

\section{BLINK: The Key Idea}
\paragraph{\textbf{Layer Score as Weight Corruption Indicator}}\label{layer_score}
During inference, a BN layer normalizes activations using fixed running statistics ($\mu_{running}$, $\sigma_{running}$) and affine  parameters ($\gamma$ and $\beta$), all frozen after training.
\begin{equation}\label{eqn:bn_eq}
\begin{aligned}
\hat{x} &= \frac{x - \mu_{\mathrm{running}}}{\sigma_{\mathrm{running}}}, \qquad y = \gamma\hat{x} + \beta,\\[4pt]
\end{aligned}
\end{equation}
Here, $x$ and $y$ are BN inputs and outputs, respectively. The core foundation of BLINK is that the frozen running statistics serve as a per-channel reference for healthy activations. An upstream weight fault pulls the inference-time batch statistics away from this reference, leaving a deviation that we quantify per channel and average into a layer score ($S_\ell$) with Eqn. \ref{eqn:layer_score}.
\begin{equation}\label{eqn:layer_score}
\begin{aligned}
z_{\mu}^{l,c} &= \frac{\mu_{b}^{l,c} - \mu_{\mathrm{running}}^{l,c}}{\sigma_{\mathrm{running}}^{l,c}},
\qquad
S_l = \frac{1}{C}\sum_{c=1}^{C} \big|z_{\mu}^{l,c}\big|.
\end{aligned}
\end{equation}
Here, $\mu_{b}^{l,c}$ is the inference-time batch mean of channel $c$ at layer $l$ and $C$ is the channel count.

\paragraph{\textbf{Overview of BLINK}}
Fig. \ref{fig:overview} explains BLINK at a glance, illustrating the two phases: (1) offline pre-characterization and (2) on-chip detection and mitigation. BLINK rests on a key observation: weight corruptions, especially the ones likely to impair accuracy,  do not stay hidden in the weights but shift and scale the activation distribution. The larger this shift, the larger the eventual accuracy degradation is likely to be (Fig.~\ref{fig:overview}(g)). BLINK quantifies the shift at each BN layer with the layer score $S_\ell$ (see ~\eqref{eqn:layer_score}) and utilizes its correlation with the end-to-end accuracy drop (Fig.~\ref{fig:overview}(b)) for runtime prediction of the expected accuracy degradation due to  accumulated perturbations. During offline characterization, the profile of accuracy drop versus $S_\ell$ is used to calibrate a detection threshold, $T_\ell$ (Fig. ~\ref{fig:overview}(b, c)). Also, the clean BN reference for each checkpoint is characterized and stored on-chip (Fig. ~\ref{fig:overview}(d)), providing the target for mitigation.

During runtime, layer score ($S_\ell$) is computed (Fig. ~\ref{fig:overview}(e)) and compared with $T_\ell$, leading to the following three scenarios: (1)~If all scores stay below their thresholds ($S_\ell<T_\ell$), activations pass downstream normally. (2)~If a checkpoint crosses its threshold ($S_\ell>T_\ell$), it passes through the BN layer with updated running statistics, where the activation distribution is recentered and rescaled to mitigate the effect of weight corruptions before sending it to the next layer (Fig. ~\ref{fig:overview}(f, g)). (3) If the residual shift after mitigation shows that the model has been pushed past its recoverability bound, a watcher aborts the inference to avert the generation of unreliable outputs.

We note two important points. First, BLINK is not limited to models with BN layers. The same concept can be applied to BN-fused architectures with $\mu_{running}$ and $\sigma_{running}$ replaced by 0 and 1, respectively, but requires a separate detection/mitigation layer. In this paper, our analyses focus on the former models. Second, if  BLINK mitigation is not sufficient (depending on the accuracy targets), BLINK can be used primarily for detection and complemented with other mitigation techniques \cite{quan2022training,bhattacharjee2023examining} based on system needs and resources (access to host, labeled data). However, if BLINK mitigation is sufficient, it offers a distinct advantage of full autonomy.

To explain BLINK's operation in detail, we start by discussing the on-chip detection and mitigation process, including the BLINK parameters and thresholds needed. With the understanding of on-chip operation, we present the offline pre-characterization procedure of the BLINK parameters, which emulates the on-chip functionality to achieve high detection/mitigation efficacy.

\section{On-chip Operation}
\label{runtime}
\paragraph{\textbf{BLINK parameters stored on-chip}} To carry out on-the-fly detection and mitigation, the accelerator needs to store the following parameters (obtained through offline characterization - see Section \ref{offline}): (1) Detection threshold for each layer $l$ ($T_\ell$), (2) Abort threshold for the watcher ($T_w$), (3) Recentering and rescaling parameters for each channel $c$ of each layer $l$ ($z_{\mu,tgt}^{l,c}$ and $z_{\sigma,tgt}^{l,c}$, respectively), and (4) trust gate bit (TGB) for each layer.

\paragraph{\textbf{Detection and Mitigation}}\label{detection_and_mitigation} During inference, as each layer produces the output activations, BLINK computes batch mean of the activation $\mu_{b}^{l,c}$ and the layer score $S_l$ based on Eqn. ~\eqref{eqn:layer_score}. In addition, the standard deviation of the activations $\sigma_{b}^{l,c}$ is computed. If $S_\ell$ > $T_\ell$, a flag signal ($flag_l$) is asserted. In this case, BLINK replaces $\mu_{running}^{l,c}$ and $\sigma_{running}^{l,c}$ with $\mu_{new}^{l,c}$ and $\sigma_{new}^{l,c}$, respectively for batch normalization, with an objective to bring the activations closer to the clean distribution. Here,
\begin{equation}\label{eqn:mu_new}
\begin{aligned}
\mu_{new}^{l,c} &= \mu_{b}^{l,c} - z_{\mu,tgt}^{l,c}\sigma_{new}^{l,c}
\qquad
\sigma_{new}^{l,c} &= \frac{\sigma_{b}^{l,c}}{z_{\sigma,tgt}^{l,c}}
\end{aligned}
\end{equation}
Note that mitigation exploits the fact that BN already stores a reference for healthy activation statistics. For each channel, we adjust the running mean to reduce the shift due to weight corruptions and rescale the running variance toward its stored target. This process is repeated at each flagged layer. Thus, BLINK re-centers the corrupted signal at its point of origin, flowing near-clean activations into the subsequent layers and preventing the corruptions from compounding downstream (Fig. ~\ref{fig:overview}(g)).

\paragraph{\textbf{Abort using a Watcher}}
\label{abort_method}
If weight corruptions are too large to be mitigated, BLINK issues an abort signal indicating that the chip is not usable. In other words, mitigation may be applied at many layers, yet the accelerator may remain too damaged to use. To enable the abort mode, we set the final BN layer as our watcher layer, since the activation coming to final BN layer will carry all the previous perturbation and mitigation information. Besides, since the abort situation is expected to be uncommon, we do not keep too many checkpoints to avoid unnecessary overhead. At the watcher layer, we compute the watcher score $S_w$ (similar to the layer score) and if $S_w$ > $T_w$, an abort signal is issued, else normal operation continues.

\paragraph{\textbf{BLINK + Other Mitigation Approaches}}\label{BLINK_&_others}
As we will discuss in Section \ref{mitigation_results}, BLINK significantly boosts the accuracy of DNNs under weight corruptions, while remaining fully autonomous. However, if the system allows trading-off autonomy to some extent (for instance, permitting communication with the host, or having access to labeled data), other mitigation approaches such as FPT \cite{quan2022training} and BN-FT \cite{bhattacharjee2023examining} can provide even better accuracy recovery under certain conditions (details later). It may be mentioned that although BN-FT in \cite{bhattacharjee2023examining} has been proposed for weight drifts in in-memory computing (IMC) architectures; here, we re-purpose it to combat a variety of weight corruptions while also extending its applicability to non-IMC platforms.  However, such techniques, on their own, still cannot provide continuous monitoring (required, for instance, to combat silent data corruption, weight drifts and other similar scenarios). In this context, we propose using BLINK primarily for detection based on continuous monitoring of harmful corruptions. For this, we utilize a trust gate bit (TGB) for each layer (more details in Section \ref{trust_gate}). TGB = 1 denotes that the detection of the layer should be trusted, else, it should be suppressed (see Fig. \ref{fig:overview}). If $flag_l$ signal of any \textit{trusted} layer is asserted, an alarm signal is raised. In this case, BLINK's mitigation can be combined with FPT or BN-FT to provide even better accuracy. For example, if drift gradually corrupts the weights, BLINK would raise an alarm once this corruption is large enough to impair accuracy. Now, the system can halt the operation, apply FPT or BN-FT and resume inference. Thus, BLINK could be utilized either as a stand-alone fully autonomous detection and mitigation framework (as discussed in the first three sub-sections), or in conjunction with other mitigation techniques  trading-off autonomy for better mitigation performance while providing continuous on-the-fly detection.

\section{Offline Profiling and BLINK Parameters}
\label{offline}
Offline pre-characterization of BLINK parameters ($T_\ell$, $T_w$, $z_{\mu,tgt}^{l,c}$ and $z_{\sigma,tgt}^{l,c}$) is an important step which determines BLINK's efficacy. This process requires detailed profiling of the dependence of accuracy drop on the layer score considering clean and corrupted models and using these trends to set the thresholds and reference parameters based on the user specifications of target accuracy recovery.

\paragraph{\textbf{Profiling Procedure and Setting Thresholds}}\label{profiling}
The profiling of layer (watcher) score with respect to the accuracy drop sets $T_l$ ($T_w$) and therefore, must emulate the on-chip operation (as discussed in the previous section). To profile layer $k$ (Fig.\ref{fig:overview}(a)), we inject weight perturbations into the sub-network from the input up to layer $l$ and sweep the fault rate. For every upstream layer that is flagged, we apply BLINK mitigation (based on recentering and rescaling - see Section \ref{detection_and_mitigation}), exactly as would be applied during the inference. We then note the residual layer score $S_l$ at the still-faulted layer $l$ and pair it with the resulting accuracy drop. Layer $l$ itself is left uncorrected, since it is the layer whose detectability we are characterizing. Repeating this across all layers and fault rates yields, per checkpoint, a set of layer score and accuracy drop observations. Because each checkpoint is read only after its upstream faults have been corrected, its score reflects the damage that originates at or propagates past that point, rather than damage that an upstream correction would already have mitigated.

Plotting these numbers gives us a layer score versus accuracy drop trend (Fig. \ref{fig:overview}(b)). For a fixed user-specified accuracy drop $\tau$ (= $10\%$ in Fig. \ref{fig:overview}(c)) and a variable layer score, 4 sub regions of this plot are obtained which yield true positives (TP), true negatives (TN), false positives (FP) and false negatives (FN). This is repeated for different layer score values and a receiver operating characteristic (ROC) curve is obtained (Fig. \ref{fig:overview}(c)). Note, the parameter $\tau$ denotes the threshold for accuracy drop above which, the corruptions are deemed harmful. Thus, the area under the ROC curve (AUC) measures how well the layer score ranks corrupted models above clean ones, independently of where any threshold is placed. The AUC is also called reliability score of that layer.

Now, using this ROC, we determine the detection threshold $T_\ell$ which corresponds to $1\%$ false-negative rate (Fig. ~\ref{fig:overview}(c)). It is important to note that the two types of errors, FP and FN (Fig. ~\ref{fig:overview}(b)) are not equally costly for a stand-alone BLINK detection and mitigation. A missed harmful fault (FN) propagates to the output leading to incorrect inference. In contrast, a false alarm (FP) only triggers mitigation, which merely induces a negligible recentering and rescaling of a healthy layer's activation distribution, causing no meaningful degradation. Thus, we constrain FNR and let FPR follow.

Note that the target for BLINK detection is not just catastrophic collapse, which produces large and easily flagged deviations. For instance, previous works in \cite{chen2021low,9643556,9474113} work well for accuracy drops as large as 50\%. In contrast, BLINK works well for subtle degradations as well, with a moderate accuracy drop (e.g. 5\% or 10\%).

Similarly, we set the abort threshold ($T_w$). The only difference is that instead of $\tau$, we use a larger accuracy drop $\tau_{abort}$ (=50\%) to obtain the ROC. Here, $\tau_{abort}$ denotes the catastrophic accuracy degradation which BLINK can easily detect but cannot meaningfully mitigate.

\paragraph{\textbf{Recentering and Rescaling Parameters}}
The recentering and rescaling parameters ($z_{\mu,tgt}^{l,c}$ and $z_{\sigma,tgt}^{l,c}$, respectively) are defined as follows.
\begin{equation}
z_{\mu,tgt}^{l,c} = \frac{\mu_{\mathrm{clean}}-\mu_{\mathrm{running}}}{\sigma_{\mathrm{running}}},
\qquad
z_{\sigma,tgt}^{l,c} = \frac{\sigma_{\mathrm{clean}}}{\sigma_{\mathrm{running}}}.
\label{eq:zmu_zsig}
\end{equation}
Here, $\mu_{\mathrm{clean}}$ and $\sigma_{\mathrm{clean}}$ are the mean and standard deviation of the activation distribution on clean model. From  ~\eqref{eq:zmu_zsig}, we get per channel $z_{\mu,tgt}^{l,c}$ and $z_{\sigma,tgt}^{l,c}$ for each batch. We run a clean model over 200 training batches of size 128, compute parameters from ~\eqref{eq:zmu_zsig} and calculate their median to obtain $z_{\mu,tgt}^{l,c}$ and $z_{\sigma,tgt}^{l,c}$. Using the median values suppresses batch-to-batch noise and bounds $z_{\sigma,tgt}^{l,c}$ to $[.5,2]$ for numerical stability.

\paragraph{\textbf{Trust Gates for Enhancing BLINK Detection}}\label{trust_gate}
As discussed before, when BLINK is used as a standalone detection/mitigation framework, FPs are not particularly harmful and can be traded off to achieve low FNR. However, when BLINK is used in conjunction with other mitigation techniques (see Section \ref{BLINK_&_others}), it primarily serves to continually monitor to detect issues on-the-fly, which invokes mitigation processes (such as FPT and BN-FT) that may be costly. In this case, reducing FPs is also important to avert unnecessary mitigation calls. To enhance the detection efficacy of BLINK, we use trust gates. For this, we run the clean model over the training set at the profiling batch size with all thresholds active, and count which layers raise an alarm. Our analysis shows that mostly the layers with smaller AUC flag unnecessarily. Hence, we find the largest AUC among those extra sensitive layers and use that AUC threshold as $AUC_{th}$ during detection. Any layer with $AUC> AUC_{th}$ is allowed to flag, while for the other layers, flagging is suppressed. This helps reduce both FN and FP.

\section{Experimental Setup}\label{sec:setup}
We evaluate ResNet-20(clean accuracy (CA)=88.95\% on CIFAR-10), ResNet-50(CA=94.35\%/78.56\% on CIFAR-10/100) and MobileNetV2(CA = 91.96\%/69.89\% on CIFAR-10/100). Weights are quantized to 8-bits in the 2's complement form following a symmetric per layer fixed point method. Activations are in FP32 format.

\paragraph{\textbf{Fault Injection Setup:}}\label{fault_injection_setup}
We evaluate BLINK under four fault models implemented by injecting bit-corruptions in the 8-bit weights.
\textbf{Global random faults} are injected uniformly across the whole accelerator considering the \textbf{stuck-at fault} (SAF) and bit-flip models. These model transient upsets (radiation induced faults, thermal noise), near-threshold voltage scaling\cite{stutz2021bit} or aging . We sweep fault rates from $0\%$ to $2\%$ and report an average over 100 independent trials with a fresh test batch and fault pattern. The safety-abort analysis uses this random-fault model at elevated rates 0.50\%-4\%.
\textbf{Spatially localized faults:} In a bit-sliced layout, each bit position of the INT8 weight is stored in a separate memory array. Considering the worst case, we pick the arrays with most significant bits (MSBs) only. We randomly select a contiguous block of size $B_{block}$ and consider that a sub-block of size $f_{SB}B_{block}$ is corrupted. Within this sub-block, we randomly flip the MSB of  fraction (c) of weight bits. Sweeping $f_{SB} \in \{0.05, 0.1, 0.2, 0.5\}$ and $c \in \{0.1, 0.2, 0.3, 0.5\}$ gives an effective damaged fraction $f_{\mathrm{eff}} = f_{\mathrm{SB}} \cdot c$ from $0.5\%$ to $25\%$.
\textbf{Progressive bit-flip attacks (PBFA)} follow the progressive bit search of~\cite{rakin2019bit}. Each iteration ranks candidate bits by gradient magnitude ($\partial L/\partial w$) and flips the single most damaging bit in each iteration. The corrupted model is re-ranked following the same process in the next iteration. We assume a non-adaptive attacker that does not jointly optimize damage and detector evasion. We run $100$ independent attacks with an attack batch size of $128$ and track accuracy over the first ten flips.

\paragraph{\textbf{BLINK + Other BN-based techniques:}}\label{baseline}
We explore the use of BLINK's detection with two previously proposed BN-based recovery methods using the configurations reported in the original works. For (FPT)~\cite{quan2022training}, we re-estimate $\mu_{running}$ and $\sigma_{running}$ using forward passes over a small, unlabeled calibration set of 32 images for 32 iteration cycles to produce the best case for the baseline~\cite{quan2022training}.  For BN-FT~\cite{bhattacharjee2023examining}, we fine-tune the BN affine parameters ($\gamma$ and $\beta$) via supervised gradient descent for $5$ epochs with 5000 training images.

\begin{figure*}[t]
    \centering
    \includegraphics[width=1\linewidth]{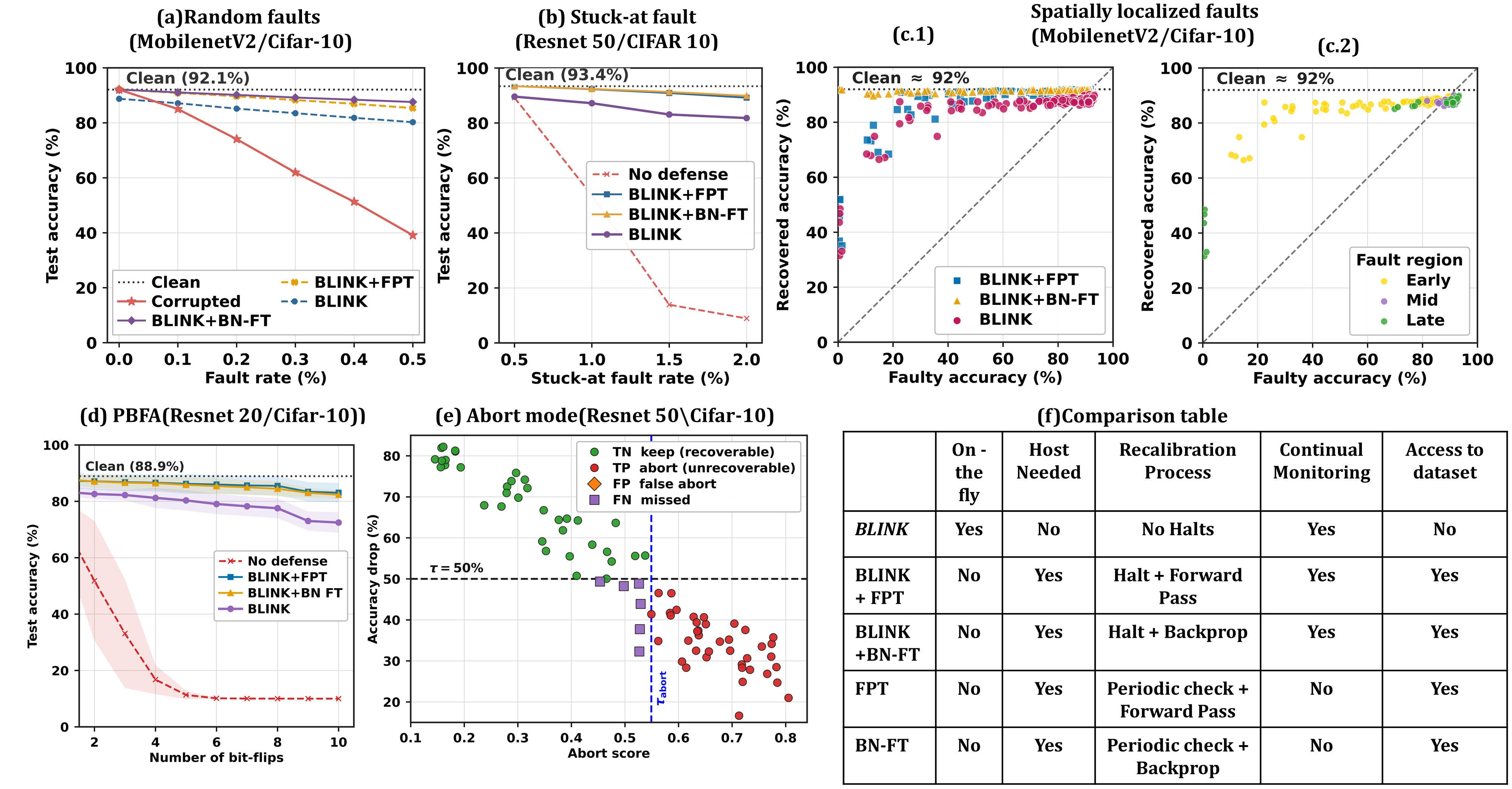}
    \caption{\footnotesize BLINK mitigation across all four fault types, stand-alone and combined with offline BN methods (BLINK+FPT, BLINK+BN-FT). (a)~Random bit flips (MobileNetV2/CIFAR-10). (b)~Stuck-at faults (ResNet-50/CIFAR-10). (c.1)~Localized faults (MobileNetV2/CIFAR-10): recovered vs.\ faulty accuracy; dashed diagonal is no recovery. (c.2)~Same, BLINK-only, colored by fault depth. (d)~PBFA (ResNet-20/CIFAR-10): accuracy vs.\ bit flips. (e)~Safety abort: accuracy drop vs.\ watcher score; $\tau_{abort}$ separates keep from abort. (f)~Qualitative comparison of all methods.}
    \label{fig:all_mitigation}
\end{figure*}

\section{Results}
Since mitigation is ingrained in the detection process in BLINK, detection is evaluated with mitigation enabled. The only difference is the use of trust gates for detection evaluation, while for BLINK mitigation analysis, trust gates are disabled.

\subsection{Detection}
For detection, a trial is declared harmful when at least one trusted layer crosses its threshold at any point during the pass. The ground truth is measured with end-to-end accuracy drop and the detection results are based on comparing BLINK detection with the ground truth. Trials are graded in three zones: (1) Clean trials, where no corruption is injected, determine the false-positive rate (FPR). (2) Trials with sufficiently high faults rates such that the accuracy drop $\geq \tau$ are deemed harmful and score the true-positive rate (TPR) and FNR. (3) Trials with lower fault rates result in accuracy drop below $\tau$ form a \emph{gray zone}. Since the detector was not calibrated for this regime, gray-zone alarms are reported separately to illustrate the detection efficacy of BLINK beyond its calibrated regime. Precision (= TP/(TP+FP)) is computed over the clean and harmful trials only. We evaluate BLINK at $\tau=10\%$ and $\tau=5\%$. The thresholds are calibrated once for $\tau=10\%$. The same thresholds are used to analyze $\tau=5\%$ to test the effectiveness of BLINK in catching milder degradations than it was calibrated for.

\textbf{False positive rate:} Since FPR depends only on clean trials, it is the same for all types of weight corruptions. Table. ~\ref{tab:detection} shows that BLINK achieves very small FPR (1.1\%) by virtue of the trust gates.

\textbf{Random Bit-Flip Faults:}
As seen from Table \ref{tab:detection}, the BLINK detector never misses a harmful case for the trials conducted (TPR = $100\%$ and FNR = 0) for both cases $\tau = 10\%$ and $5\%$ as intended by FNR constrained calibration. We can also see a high gray zone detection rate $83.5\%$ for $\tau=10\%$ and $73.6\%$ for $\tau=5\%$.

\textbf{Random Stuck-at Faults:}\label{detection_saf}
These are tested for Resnet-50/Cifar- 10 with accuracy drop of 10\% and with around 2500 trials showing 100\% TPR.

\textbf{Spatially Localized Bit-Flip Faults:}\label{detection_localized}
Amongst all corruption scenarios considered, the localized faults represent the worst case for activation shift detection. This is because it is challenging to detect faults which are concentrated in the later layers or final classifiers (FC), where few or no BN layers remain downstream to detect the activation shift. Despite that, BLINK exhibits excellent detection performance for localized faults (TPR around 98\% - see Table. \ref{tab:detection}).

\textbf{Bit Flip Attacks:}\label{detection_bfa}
For bit flip attacks, the detection is easier than all the other scenarios because of a large accuracy drop jump. The detector performance is $100\%$ TPR and 0\% FNR. A few trials fall into gray zone because the first iteration accuracy drop is very small.

\textbf{Safety Abort:}\label{detection_abort}
We evaluate the single-watcher gate (watcher at last BN) on MobileNetV2/CIFAR-10 over $150$ trials spanning global fault rates $0.5\% - 4\%$.
Treating a trial as positive when its corrected accuracy falls below $\tau_{abrt}$ = 50\%, the abort score $S_w$ separates recoverable from unrecoverable inferences with reliability score (AUC) of 0.98 shown in Fig. \ref{fig:all_mitigation}(e).  The watcher gate achieves a recall (=TP/(TP+FN)) of $94.4\%$ and a false-abort rate of $7.0\%$.

\begin{table}[t]
  \centering
  \caption{Detection of random and spatially localized bit-flip faults
  (MobileNetV2/CIFAR-10, batch $=$ 128)}
  \label{tab:detection}
  \begin{tabular}{@{}ll cc@{}}
    \toprule
    \textbf{Corruption} & \textbf{Metric} & $\tau{=}10\%$ & $\tau{=}5\%$ \\
    \midrule
    \multicolumn{2}{@{}l}{Clean (no fault) FPR} & 1.1 & 1.1 \\
    \midrule
    \multirow{2}{*}{Random/PBFA} & Harmful (TPR) & \textbf{100} & \textbf{100} \\
                               & Gray zone detection & 83.5 & 73.6 \\
    \midrule
    \multicolumn{2}{@{}l}{Precision} & 99.6 & 99.8 \\
    \midrule
    \multirow{2}{*}{Localized} & Harmful (TPR) & 98.5 & 96.6 \\
                               & Grey zone detection & 39.7 & 34.0 \\
    \midrule
    \multicolumn{2}{@{}l}{Precision} & 99.8 & 99.9 \\
    \bottomrule
  \end{tabular}
\end{table}

\subsection{Mitigation}\label{mitigation_results}
In this section, we will compare the mitigated accuracies for  three methods. (i) standalone BLINK (ii) BLINK+ FPT (iii) BLINK+ BN-FT as described in Section. \ref{BLINK_&_others}. All the mitigated accuracy numbers are averaged over several trials with multiple ($\sim 100$) fresh batches and fault patterns.
Fig. \ref{fig:all_mitigation} shows that BLINK recovers substantial accuracy entirely on the fly with no data access or host intervention.

For \textbf{random bit-flip faults}, (Fig. \ref{fig:all_mitigation}(a), MobilenetV2/Cifar-10), BLINK recovers accuracy from $4\% - 86\%$ for different fault rates, only with the help of profiled parameters. For different workloads and with varying fault rates, accuracy recovery achieved by BLINK is shown in Table. \ref{tab:faulty_corrected_per_fault}.  Similarly, for \textbf{random stuck-at faults} (Fig. \ref{fig:all_mitigation}(b)), BLINK's mitigation achieves excellent recovery even for very large fault rates. For \textbf{spatially localized faults} in Fig. \ref{fig:all_mitigation}(c.2), recoveries depend on fault depth. For early (yellow) and mid (purple) region, we observe a substantial accuracy recovery (data points far above the zero-recovery line). Even for some late-region (green) trials, mild recovery is observed. However, for some late region trials, recovery is not seen because few or no BN layers remain downstream to mitigate the damage. In particular, the final classifier has no BN layer after it and is therefore outside BLINK's protection, which is a limitation of our approach.

For \textbf{PBFA} ~\cite{rakin2019bit} we evaluate ResNet-20/CIFAR-10. The results are shown in Fig.~\ref{fig:all_mitigation}(d) where solid lines are the mean and shaded bands are $\pm 1$ standard deviation. Without any defense (red dashed), the attack is devastating, and a handful of flips drives accuracy from the clean $88.9\%$ down to $10\%$. The wide red band shows that the outcome is also highly erratic from trial to trial. BLINK stops the collapse and recovers accuracy significantly.

Across all fault types, method (ii) and (iii) recovers a few points more than standalone BLINK, at the cost of halting inference and using host access or using data access (Fig. \ref{fig:all_mitigation}(f)) . The gap is largest for the most severe localized faults (Fig.~\ref{fig:all_mitigation}(c)), where method (iii) recovers nearly to clean accuracy in the regime where BLINK and method (ii) do not perform that well. Labeled data and back-propagation buy that recovery for BLINK+BN-FT.

\begin{table}[t]
\centering
\small
\setlength{\tabcolsep}{3pt}
\caption{Top-1 accuracy (\%) under random bit-flips for different fault rates. FA: faulty accuracy; RA: Recovered accuracy (BLINK).}
\label{tab:faulty_corrected_per_fault}
\resizebox{\columnwidth}{!}{%
\begin{tabular}{ll cc cc cc}
\toprule
\textbf{Network} & \textbf{Dataset} & \multicolumn{2}{c}{\textbf{0.10\%}} & \multicolumn{2}{c}{\textbf{0.20\%}} & \multicolumn{2}{c}{\textbf{0.50\%}} \\
& & \textbf{FA} & \textbf{RA} & \textbf{FA} & \textbf{RA} & \textbf{FA} & \textbf{RA} \\
\midrule
ResNet-50    & CIFAR-10  & 90.14 & 91.19 & 14.48 & 84.55 & 10.00 & 85.88 \\
ResNet-50    & CIFAR-100 & 70.80 & 73.84 & 54.24 & 69.64 & \phantom{0}1.00 & 62.65 \\
\midrule
MobileNetV2  & CIFAR-10  & 85.83 & 88.78 & 76.76 & 86.80 & 47.60 & 82.20 \\
MobileNetV2  & CIFAR-100 & 41.77 & 61.73 & 35.37 & 58.80 & \phantom{0}7.84 & 52.01 \\
\bottomrule
\end{tabular}%
}
\end{table}

\section{Hardware Overhead}
We estimate the hardware overhead considering a $128 \times 128$ weight stationary systolic array structure with the assumption of $100\%$ array utilization with a functional unit of 128 ops/cycle to compute the extra computation. Inside the functional unit, we assume that the cost of computing one addition, multiplication or comparison is equal to one multiply-accumulate (MAC). Note that this leads to conservative overheads of BLINK. For full inference, several arrays are utilized to perform the per layer convolution operations.

For BLINK, the calculation of parameters,  $\mu_b$ and $\sigma_b$ lead to computational overhead but no additional latency cycles because they are computed in parallel as the activations are generated. The computation of the comparison with threshold and the BN recalibration parameters ($\mu_{new}$ and $\sigma_{new}$) cost extra cycles as well as computation. The computation overhead is calculated based on the number of MAC/addition/multiplication/comparison operations and the latency overhead is obtained based on the number of  cycles needed for each operation.
Based on this estimation model, the computation overhead of BLINK for batch size of 128 is $0.53\%$ for Resnet-50 and $4.5\%$ for MobilenetV2 for the worst-case scenario (in which each layer is flagged and mitigated). The corresponding latency overhead is $0.03\%$ for Resnet-50 and $1.05\%$ for MobilenetV2.

\section{Conclusion}
We present BLINK, an autonomous on-chip BN based in-situ detection and mitigation technique for a broad spectrum of weight corruptions. BLINK is based on monitoring the shifts in the activation statistics due to weight perturbations and recentering them upon detection of harmful corruptions. BLINK achieves excellent detection efficacy across various fault types and PBFA, and recovers significant amount of accuracy on-the-fly depending only on offline profiling and pre-characterization of parameters, leading to a fully autonomous approach. Hardware overhead estimation shows a negligible computational and latency costs incurred by BLINK. Combining BLINK detection with prior recovery methods brings accuracy even closer to the ideal value but at the cost of autonomy.

\section*{Acknowledgment}
This work was supported in part by the Center for the Co-Design of Cognitive Systems (CoCoSys), one of seven centers in JUMP 2.0, a Semiconductor Research Corporation (SRC) program sponsored by DARPA.

\bibliographystyle{plain}
\bibliography{mybib}

\end{document}